\documentclass[numbers]{sigplanconf}

\usepackage{amsmath}

\usepackage{graphicx}
\usepackage{epstopdf}
\usepackage{subcaption}
\usepackage{url}
\usepackage{tabularx}
\usepackage{enumitem}

\DeclareGraphicsExtensions{.pdf,.eps,.png}

\begin{document}

\setlength{\pdfpageheight}{\paperheight}
\setlength{\pdfpagewidth}{\paperwidth}

\conferenceinfo{BPOE '16}{April 3, 2016, Atlanta, GA, USA}
\copyrightyear{2016}
\copyrightdata{978-1-nnnn-nnnn-n/yy/mm}
\copyrightdoi{nnnnnnn.nnnnnnn}

\publicationrights{author-pays}

\titlebanner{BPOE-7 Submission (Paper \#2)}        

\title{When to use 3D Die-Stacked Memory for Bandwidth-Constrained Big Data Workloads}

\authorinfo{Jason Lowe-Power \and Mark D. Hill \and David A. Wood}
           {University of Wisconsin--Madison}
           {powerjg,markhill,david@cs.wisc.edu}

\maketitle

\begin{abstract}
Response time requirements for big data processing systems are shrinking.
To meet this strict response time requirement, many big data systems store all or most of their data in main memory to reduce the access latency.
Main memory capacities have grown, and systems with 2~TB of main memory capacity available today.
However, the rate at which processors can access this data---the memory bandwidth---has not grown at the same rate.
In fact, some of these big-memory systems can access less than 10\% of their main memory capacity in one second (billions of processor cycles).

3D die-stacking is one promising solution to this bandwidth problem, and industry is investing significantly in 3D die-stacking.
We use a simple back-of-the-envelope-style model to characterize if and when the 3D die-stacked architecture is more cost-effective than current architectures for in-memory big data workloads.
We find that die-stacking has much higher performance than current systems (up to $256\times$ lower response times), and it does not require expensive memory over provisioning to meet real-time (10~ms) response time service-level agreements.
However, the power requirements of the die-stacked systems are significantly higher (up to $50\times$) than current systems, and its memory capacity is lower in many cases.
Even in this limited case study, we find 3D die-stacking is not a panacea.
Today, die-stacking is the most cost-effective solution for strict SLAs and by reducing the power of the compute chip and increasing memory densities die-stacking can be cost-effective under other constraints in the future.
\end{abstract}




\section{Introduction}

New data is generated at an alarming rate: as much as 2.6 exabytes are created each day~\cite{Mcafee2012}.
For instance, with the growth of internet-connected devices, there is an explosion of data from sensors on these devices.
The ability to analyze this data is important to many different industries from automotive to health care~\cite{Oracle2013}.
Users want to run complex queries on this abundance of data, sometimes with real-time constraints.
These trends of increasing data, increasing query complexity, and increasing performance constraints put a great strain on our computational systems.

To meet the latency requirements of real-time queries, many big data systems keep all of their data in main memory.
In-memory queries provide lower response times than queries that access disk or flash storage, and are increasingly popular~\cite{Zhang2015}.
The in-memory data may be the entire data corpus, or the hot parts of the data set.

Many big data workloads are memory intensive, performing a small number of computational operations per byte of data loaded from memory.
This low computational intensity means these workloads are increasingly constrained by memory bandwidth, not the capabilities of the processor.
This problem is exacerbated by the increasing gap between processor speed and memory bandwidth~\cite{Burger1996}.

As a solution to this memory bandwidth gap, industry is currently investing heavily in 3D die-stacked memory technologies~\cite{AMD, NVIDIA2015, Black2013, Pawlowski2011, Samsung2014, Kim2014, IBM2011, Ranganathan2011, Casper2011, OConnor2014, Saito2012}.
3D die-stacked memory enables DRAM chips stacked directly on top of compute chip with through-silicon vias (TSVs).
TSVs enable high-bandwidth low-power memory accesses.
Currently only high-performance graphics processors use die-stacked memory.
However, many major companies such as IBM, Intel, Samsung, Micron, 3M and others are making significant investments in 3D die-stacking.

\begin{figure}
 \begin{center}
  \includegraphics[width=0.9\columnwidth]{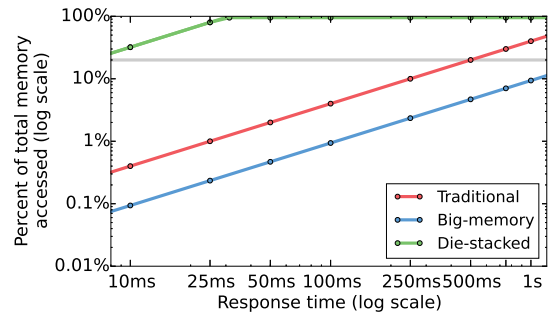}
  \caption{Comparison of the memory bandwidth-capacity tradeoff in current commercial systems.}
  \label{fig:memorywall}
  \end{center}
  \vspace{-2em}
\end{figure}

\begin{figure*}
 \begin{center}
   \begin{tabular}{l | c | c | c}
    & \Large{Traditional} & \Large{Big-memory} & \Large{Die-stacked} \\ \hline \hline
    &
    \begin{subfigure}[c]{0.25\linewidth}
     \centering
     \includegraphics[width=.75\textwidth]{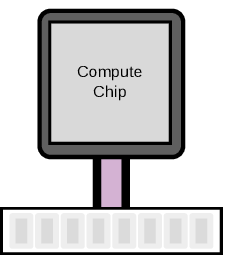}
     \caption{A traditional server memory configuration. In this configuration, DRAM DIMMs are directly connected to the compute chip. This is a commodity system available today that is configured for the maximum bandwidth-capacity ratio.}
     \label{fig:traditional-system}
    \end{subfigure}
    &
    \begin{subfigure}[c]{0.25\linewidth}
     \centering
     \includegraphics[width=\textwidth]{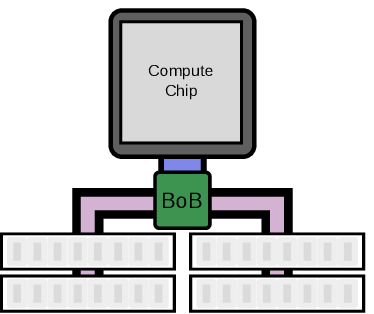}
     \caption{Big-memory server memory configuration. In this configuration, DRAM DIMMs are connected to a buffer-on-board (BoB), and the BoB is connected to the compute chip. This system is available today and is configured for the maximum potential DRAM capacity.}
     \label{fig:big-memory-system}
    \end{subfigure}
    &
    \begin{subfigure}[c]{0.25\linewidth}
     \vspace{4.9em}
     \centering
     \includegraphics[width=\textwidth]{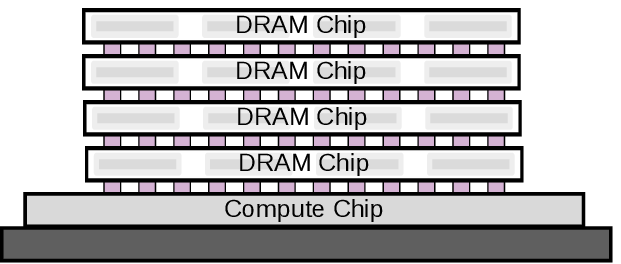}
     \caption{A 3D die-stacked server configuration. In this configuration, DRAM chips are integrated with the compute chip with through-silicon vias. This system is not yet available. It has a very high bandwidth-capacity ratio, and is tailored to bandwidth-bound in-memory workloads.}
     \label{fig:die-stacked-system}
    \end{subfigure} \\ \hline
    Off-socket bandwidth & 102 GB/s & 192 GB/s & 256 GB/s \\ \hline
    Memory per socket & 256 GB & 2 TB & 8 GB \\ \hline
    Sockets per blade & 4 & 1 & 9 \\ \hline
    Blades needed for & 16 & 8 & 228 \\
    16~TB capacity & & & \\ \hline
    Aggregate bandwidth & 6.4 TB/s & 1.5 TB/s & 512 TB/s \\
    for cluster & & & \\ \hline
    Commercial example & PowerEdge R930 \cite{Dell2016} & Oracle M7 \cite{Oracle2015} & N/A \\ \hline
   \end{tabular}
  \end{center}
  \caption{Diagrams of the three system designs we consider in this paper and potential configurations for each system.}
  \label{fig:systems}
\end{figure*}

With these investments in 3D die-stacking, it is prudent to evaluate the impact this technology may have on important workloads.
In this paper, we investigate using a 3D die-stacked system for real-time big data queries.
As a case study, we focus on an analytic database workload.
Due to recent advances in analytic database query algorithms~\cite{Li2013,Li2014,Power2015,Feng2015}, the performance of analytic database queries is now constrained by the system's main memory bandwidth~\cite{Power2015a}.
We find that the scan operator (a major contributor to the total time in analytic queries~\cite{Li2013}) requires fetching about 4 bytes to the main memory per instruction when using SIMD.
Thus, a 16 core CPU running at 3.5~GHz can theoretically generate over 200 GB/s of main-memory bandwidth!
Other, more specialized, architectures can theoretically access memory at a much higher rate.
We focus on analytic database workloads because they are one example of a big data workload that is growing in importance, has increasing datasets, and is increasingly performance constrained.

We evaluate three different server architectures, two that are available from system providers today, and the potential future architecture based on 3D die-stacking.
First, the \emph{traditional server} represents a large-memory system assembled from commodity components.
We use a Dell PowerEdge R930 as our example traditional server~\cite{Dell2016}, configured with the maximum memory bandwidth per processor.
Second, the \emph{big-memory server} represents a specialized system designed to support a very large amount of main memory.
This system is a database machine designed for in-memory database systems.
We use an Oracle M7 as our example big-memory server~\cite{Oracle2015}.
Finally, the \emph{die-stacked server} represents a novel big data machine design built around many memory-compute stacks~\cite{Black2006,Kgil2006}.
Section~\ref{sec:designs} and Figure~\ref{fig:systems} contains details of these three systems.
Throughout this work, we assume these system architectures are deployed in a cluster of computing systems.

To demonstrate the memory bandwidth gap, Figure~\ref{fig:memorywall} shows the amount of time it takes to read a fraction of the total memory capacity on a log-log scale.
The amount of time, on the x-axis, represents the performance service-level agreement (SLA) between the service provider and the user.
For instance, if the provider promises ``real-time'' access to the data (e.g., the data is queried to provide information to render a web-page) the required latency may be as little as 10--20ms.
Since we focus on bandwidth-bound workloads, we use the amount of data accessed (read or written) by the query to measure its complexity (y-axis).
In Figure~\ref{fig:memorywall}, we show this complexity as the percent of the entire main memory capacity that is accessed on each query.
As a point of reference, Figure~\ref{fig:memorywall} shows a line at 20\% of memory capacity, which is the amount of an analytic database that is referenced in a single query in a popular analytic benchmark~\cite{TPC2011}.

Figure~\ref{fig:memorywall} shows that the die-stacked design is better than the traditional and big-memory designs for low-latency bandwidth-limited workloads.
The reason for this benefit is the die-stacked system has a much higher bandwidth-capacity ratio (memory bandwidth per byte of memory capacity): 80--341$\times$.
This performance benefit exists even though the off-socket bandwidth is only 1.3--2.5$\times$ higher than the other systems because the capacity per socket is much smaller for the die-stacked system (32--256$\times$).
If a workload is bandwidth-constrained a higher bandwidth-capacity ratio will increase its performance.
For instance, to access 20\% of the total memory capacity, the big-memory server takes over 2 seconds, the traditional server takes 500~ms, and the die-stacked server takes less than 10~ms, $50\times$ faster than the current systems.

Although one could conclude from Figure~\ref{fig:memorywall} that the die-stacked system is always better than the current systems, we find that the die-stacked system \emph{may be better} under some constraints.
We use a simple back-of-the-envelope-style model to characterize \emph{when} the die-stacked architecture beats the current architectures for in-memory big data workloads.
We investigate designing cluster systems under three different system design constraints: performance provisioning (the cluster is required to meet a certain SLA), power provisioning (the cluster has a certain power budget), and data capacity provisioning (the cluster has a set DRAM capacity).

We find:
\vspace{.5em}
\begin{itemize}[nosep]
 \item When provisioning for performance, die-stacking is preferred only under aggressive SLAs (e.g., 10~ms) and the traditional and big-memory systems require memory over provisioning (up to 213$\times$ more memory than the workload requires).
 \item When provisioning for power, die-stacking is preferred with large power budgets and the big-memory system provides the most memory capacity.
 \item When provisioning for capacity, die-stacking reduces response time by up to $256\times$ and uses less energy, but die-stacking uses more power (up to $50\times$).
\end{itemize}


\section{Background}
\label{sec:background}

There are new algorithms for analytic database queries that convert complex queries into simple operations like scans and aggregates~\cite{Li2014}.
Additionally, new hardware accelerators, like general-purpose graphics processing units (GPGPUs) show potential to be both higher performance and lower energy than traditional multicore CPUs for these simpler algorithms~\cite{Power2015,Power2015a}.
Emerging hardware like GPGPUs are more efficient than CPUs because they have less complex hardware than multicore CPUs, which works together with the simplified analytic database query algorithms to increase performance and decrease energy.

Converting complex database queries into many simple operations can increase database performance.
As an example, WideTable~\cite{Li2014} converts most analytic database queries to simple scan and aggregate operations.
Once these queries are converted into more simple operations, optimized scan algorithms~\cite{Chen2001,Johnson2008} including BitWeaving~\cite{Li2013}, BitWarp~\cite{Power2015}, and SIMD-Scan~\cite{Willhalm2009} and aggregate algorithms~\cite{Feng2015} can be used.
These algorithms use the compute cores (CPU or GPU) more efficiently, which puts a strain on the memory system to keep the cores supplied with data.

As the algorithmic approaches to analytic database operations move towards more simple software operators, hardware design is supporting higher performance of simple operators.
For instance, single-instruction multiple-data (SIMD) hardware on commodity chips supports more bits per operation than ever before, up to 512-bits in AVX-512~\cite{Intel2015}.

In addition to CPUs supporting wider vector instructions, general-purpose GPUs (GPGPUs) which support very wide vectors (up a logical 4096-bit vector) are becoming more programmable.
GPUs have an order of magnitude more execution units then CPUs and each of these execution units is simpler and thus more energy efficient than their CPU counterparts.
Because of this increase in execution units GPGPUs can be higher performance and lower energy than the CPU for analytic database applications~\cite{Power2015,He2014}.

These two trends fundamentally change the nature of in-memory analytic database workloads.
Systems today store the entire database or at least the hot data in main memory.
This decreases the latency and increases the bandwidth to access the data compared to storing it on disk.
However, we have reached a ceiling in the performance of the computational cores (e.g., processing more than one record per cycle~\cite{Li2013}).
Now, the memory bandwidth is the constraint, not the computational hardware.
Now, the main performance impediment is data movement from memory to the processor.

Another exciting trend in computer architecture is 3D die-stacking.
3D die-stacking has two main characteristics that benefit analytic database workloads.
First, die-stacking allows designers to tightly couple chips with disparate technologies.
For instance, a package can contain DRAM stacked on top of compute chips.
In addition, it is also possible to stack multiple compute chips with one another (e.g., a CPU and a GPU).
Second, die-stacking can significantly increase the bandwidth to main memory, if the memory is stacked on top of the compute chip.
With die-stacking the interface to memory is much wider and can be clocked faster due to shorter wire length leading to higher memory bandwidth.
High bandwidth memory is one current die-stacked technology with 256~GB/s of bandwidth per stack~\cite{JEDEC2015}.

Although 3D die-stacking is not currently available except in the graphics market~\cite{NVIDIA2015,AMD}, many companies are making significant investments into 3D die-stacking~\cite{Black2013}.
These companies include DRAM manufacturers (Micron~\cite{Pawlowski2011}, Samsung~\cite{Samsung2014}, SK hynix~\cite{Kim2014}), system designers (IBM~\cite{IBM2011}, HP~\cite{Ranganathan2011}), silicon providers (AMD~\cite{AMD}, Intel~\cite{Casper2011}, NVIDIA~\cite{OConnor2014}), and materials manufacturers (3M~\cite{Saito2012}) as evidenced by a variety of white papers and recent press releases.

\section{Potential big data machine designs}
\label{sec:designs}

Figure~\ref{fig:systems} shows diagrams of the three different server architectures we investigate in this paper: a traditional server, a big-memory server, and a die-stacked server.
The traditional server and big-memory server are systems that can be purchased today, and
the die-stacked system is a future-looking machine architecture proposal.
When evaluating these server designs, we assume many servers are networked together into a larger cluster.

The traditional server, shown in Figure~\ref{fig:traditional-system}, is based on a commodity Intel Xeon platform~\cite{Intel}.
The main memory is accessed directly from the memory controller on the compute chip and there are four 25.6~GB/s memory controllers on the chip for a total of 102~GB/s of off-chip memory bandwidth.
Each memory controller can have up to 6 DRAM DIMMs connected.
However, since our workload is bandwidth bound, we evaluate a system configured for the minimum capacity while still achieving the maximum bandwidth.
Thus, we only populate each channel with two DDR4 DIMMs.
Each DDR4 DIMM can have 32~GB of capacity, for a total of 256~GB of main memory capacity per socket.
A similar system to what we evaluate is available from OEMs including Dell (e.g., the PowerEdge R930~\cite{Dell2016}).

The big-memory system (Figure~\ref{fig:big-memory-system}) is a proxy for database appliance solutions from companies such as Oracle and IBM.
Rather than directly access main memory like the traditional server, these systems use buffer-on-board chips to increase the maximum memory capacity.
These buffer-on-board chips also increase the peak off-chip bandwidth by using different signaling protocols than DRAM DIMMs.
Each buffer-on-board communicates both with the compute chip via a proprietary interconnect and up to eight DRAM DIMMs via the DDR4 standard.
We use the Oracle M7~\cite{Oracle2015} as an example big-memory system.
The M7 has eight buffer-on-board controllers per socket, for a total memory capacity of 512~GB and 192~GB/s per socket.

Finally, we also evaluate a novel architecture which combines compute chips with tightly-integrated memory via 3D die-stacking as shown in Figure~\ref{fig:die-stacked-system}.
This design is similar to the nanostore architecture~\cite{Ranganathan2011}.
Die-stacking technology increases the potential off-chip bandwidth by significantly increasing the number of wires between the chip and the memory.
There are currently no systems which use this technology in the database domain.
However, this technology is gaining a foothold in the high-performance graphics community~\cite{AMD,NVIDIA2015}.
We assume a system that uses the high bandwidth memory (HBM) 2.0 standard~\cite{JEDEC2015,OConnor2014}.
With HBM 2.0, each stack can be up to eight DRAM chips high and each chip has 8 gigabits of capacity, for a total capacity of 8~GB per stack.

For all of the platforms, we assume the same DRAM chip technology: 8 gigabit chips.
This gives a constant comparison in terms of capacity and power.
Additionally, assuming 3D die-stacked DRAM reaches commodity production levels, the cost-per-byte of 3D DRAM will be about the same as the cost of DRAM DIMMs.

\section{Methodology: Simple analytical model}
\label{sec:model}

Since it would be cost prohibitive to actually build and test these systems, we use a simple analytical model to predict the performance of analytic database queries on each system.
This model uses data from two sources.
First, we use data from manufacturers as in Table~\ref{fig:systems}, and
second, we assume that the query is constrained either by the memory bandwidth or the rate at which instructions can be issued by the processor.

To calculate the compute performance and power, we performed experiments on a modern hardware platform.
We used the algorithms from Power et. al~\cite{Power2015} and ran the tests on an AMD GPU~\cite{AMD2012}.
We chose to perform our tests using a GPU as it is currently the highest performing and most energy efficient hardware platform for scans~\cite{Power2015}.
We found that each GPU core uses about 3 watts when performing the scan computation, and each core can process about 6~GB/s of data.
Finally, we assume that each GPU chip has a maximum of 32 cores.

Equations~\ref{equ:modules}--\ref{equ:power} show the details of our simple model.
The inputs to our model, are shown in Table~\ref{tab:inputs}.
All of the inputs come from datasheets or manufacturer information~\cite{Dell2016,Intel,Oracle2015,JEDEC2015,OConnor2014}.
A \emph{memory module} refers to the minimum amount of memory that can be added or removed.
In the  traditional system this refers to a DRAM DIMM.
In the big-memory system a buffer-on-board and its associated memory is considered a module.
Finally, in the die-stacked system, each stack of DRAM is considered a module.
The \emph{memory channel} is how many memory modules can be connected to a single compute chip.

\begin{equation} \label{equ:modules}
\text{mem modules} = \frac{\text{db size}}{\text{module capacity}}
\end{equation}
\begin{equation} \label{equ:chips}
\text{compute chips} = \left\lceil \frac{\text{mem modules}}{\text{mem channels}} \times \frac{1}{\text{channel modules}} \right\rceil
\end{equation}
\begin{equation} \label{eq:bandwidth}
\text{chip bandwidth} = \text{mem channels} \times \text{channel bandwidth}
\end{equation}
\begin{equation} \label{equ:perf}
\text{chip perf} = min \left\{ \text{core perf} \times \text{chip cores}, \text{chip bandwidth} \right\}
\end{equation}
\begin{equation} \label{equ:cores}
\text{chip cores} = \left\lceil \frac{\text{chip perf}}{\text{core perf}} \right\rceil
\end{equation}
\begin{equation} \label{equ:mem-power}
\text{mem power} = \text{mem modules} \times \text{module power}
\end{equation}
\begin{equation} \label{equ:compute-power}
\text{compute power} = \text{chip cores} \times \text{core power} \times \text{compute chips}
\end{equation}
\begin{equation} \label{equ:blades}
\text{blades} = \left\lceil \frac{\text{compute chips}}{\text{blade chips}} \right\rceil
\end{equation}
\begin{equation} \label{equ:response-time}
\text{response time} = \frac{\text{percent accessed} \times \text{db size}}{\text{chip perf} \times \text{compute chips}}
\end{equation}
\begin{equation} \label{equ:power}
\text{power} = \text{mem power} + \text{compute power} + \text{blades} \times \text{blade power}
\end{equation}

There are three inputs to our model that are workload dependent.
The \emph{core perf} which represents the rate the core can process data, the database size (\emph{db size}) which represents the required memory capacity, and the \emph{percent accessed} which we use as a proxy for query complexity.
The percent accessed is the amount of data that a single query or operation over the data touches.
The performance estimation is based on the performance of the scan operation, but other database operations in analytic databases can be performed with similar algorithms as the scan operator~\cite{Feng2015}.

To investigate the tradeoffs between the three systems, we chose a workload size of 16~TB and a complexity of 20\% of that data, or 3.2~TB.
We believe 16~TB is a reasonable size for a large analytic database and there are some analytic queries that access 20\% of the entire database corpus~\cite{TPC2011}.
Although it is unlikely that a single operation will touch 3.2~TB of data, each query is made up of many operations, that together could touch many of the database columns.
Additionally, in the future, the data size will be increasing along with the query complexity.

In Section~\ref{sec:results} we evaluate the systems with our model under three different conditions, constant response time, constant power, and constant memory capacity.
Equations~\ref{equ:modules}--\ref{equ:power} show our model for a constant memory capacity.
We modify this slightly when holding other characteristics constant.
For constant response time, we assume an increased number of sockets to support the required bandwidth.
For constant power, we first assume each blade is fully populated, then compute the total blades that can be deployed given the power budget.

The model we use is released online as an IPython notebook at \url{https://research.cs.wisc.edu/multifacet/bpoe16_3d_bandwidth_model/}.

\begin{table}
 \begin{subtable}{\columnwidth}
 \centering
\begin{tabular}{|l||c|c|c|}
\hline
\multicolumn{1}{|c|}{}                                             & Traditional    & Big-memory & Future      \\
\multicolumn{1}{|c|}{}                                             & server         & server     & die-stacked \\ \hline \hline
\begin{tabular}[c]{@{}l@{}}module\\ capacity\end{tabular}      & 32 GB             & 512 GB       & 8 GB          \\ \hline
\begin{tabular}[c]{@{}l@{}}channel\\ bandwidth\end{tabular} & 25.6 GB/s          & 48 GB/s        & 256 GB/s        \\ \hline
\begin{tabular}[c]{@{}l@{}}memory\\ channels\end{tabular}          & 4              & 4          & 1           \\ \hline
\begin{tabular}[c]{@{}l@{}}channel\\ modules\end{tabular}          & 2\footnotemark & 1          & 1           \\ \hline
\begin{tabular}[c]{@{}l@{}}module\\ power \end{tabular}         & 8 W             & 100 W       & 10 W         \\ \hline
\begin{tabular}[c]{@{}l@{}}blade\\ chips\end{tabular}              & 4              & 1          & 9           \\ \hline
\end{tabular}
 \vspace{1em}
 \end{subtable}
 \begin{subtable}{\columnwidth}
 \centering
 \begin{tabular}[t]{| c | c | c | c | c |}
   \hline
   \multicolumn{5}{| c |}{Shared inputs} \\ \hline
    & & & & percent \\
   core perf & core power & chip cores & db size & accessed \\ \hline \hline

   6 GB/s & 3 W & 32 & 16 TB & 20\% \\ \hline
 \end{tabular}
 \end{subtable}
 \caption{Inputs for the analytical model.}
 \label{tab:inputs}
 \vspace{-1em}
\end{table}

\footnotetext{The traditional server can have up to six DRAM DIMMs per channel, but must have at least two to take advantage of the full DDR bandwidth. Therefore, we use two as the modules per channel to maximize the memory bandwidth-capacity ratio.}

\section{Results}
\label{sec:results}

\begin{figure*}[t]
 \begin{minipage}{.67\textwidth}
  \includegraphics[width=4.5in]{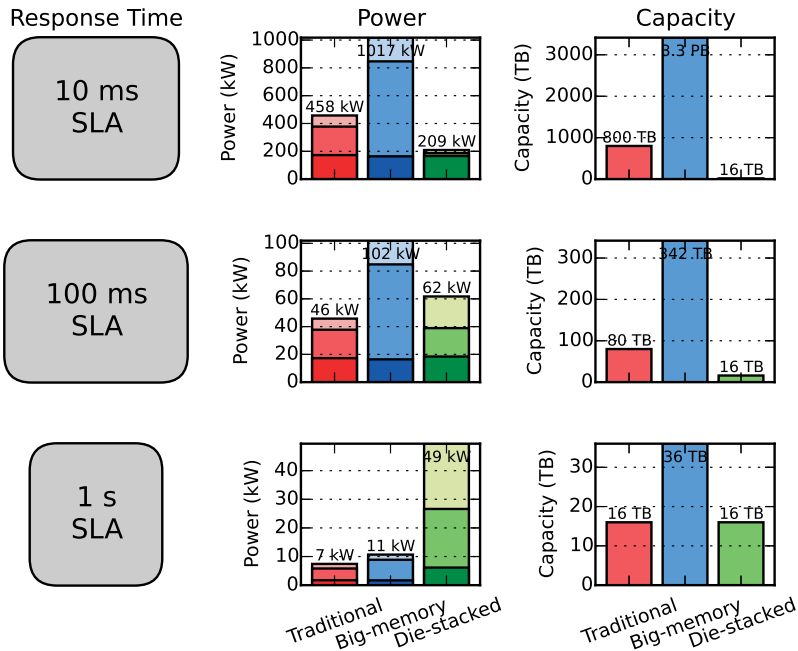}
 \end{minipage}
 \begin{minipage}{.33\textwidth}

  \begin{center}
  \fontsize{10pt}{10pt}\selectfont
  \underline{Performance Provisioning Takeaways}
  \end{center}

  \textbf{Die-stacking is preferred only under aggressive SLAs.}
  \begin{itemize}
   \item Die-stacking uses 2--5$\times$ less power and does not need memory capacity over provisioning.
   \item The crossover point is 60~ms, above which the traditional system uses less power.
  \end{itemize}
  \vspace{1em}

  \textbf{The traditional and big-memory systems require memory over provisioning.}
  \begin{itemize}
   \item Non die-stacked systems require 50--213$\times$ more DRAM than the workload.
   \item The traditional system does not require over provisioning under relaxed SLAs.
  \end{itemize}

 \end{minipage}

 \caption{Power and memory capacity of a cluster based on traditional, big-memory, and die-stacked servers with constant response times of either 10~ms, 100~ms, or one second (1000~ms). The power for each is broken down into overhead power (light, top), memory power (middle), and compute power (bottom, dark). Memory capacity must be over provisioned (16~TB required database size) to achieve the required bandwidth to meet the SLA for the traditional and big-memory systems.}
 \label{fig:constant-perf}
\end{figure*}

In this section, we investigate the tradeoffs between performance and cost in designing a real-time big data cluster using the model described in Section~\ref{sec:model}.
We consider the three designs discussed in Section~\ref{sec:designs}.
Our performance metric is the response time (i.e., the latency to complete the operation).
We chose latency instead of the throughput for two reasons.
First, in big data workloads, the latency to complete operations is increasingly important with many applications requiring ``real-time'' or ``interactive'' results.
Second, a lower latency design implies the design has higher throughput.
With no parallelism between queries, throughput is equivalent to the inverse of latency.
However, higher throughput does not always imply lower latency due to parallelism.

Cost is more difficult to quantify than performance.
Rather than try to distill cost into a single metric, we look at two important factors in the cost of building a large scale cluster.
We explore the power consumed for the cluster.
Often, power provisioning is one limiting factor in datacenter design~\cite{Barroso2013}.
Additionally, energy is a significant component of the total cost of ownership of a large-scale cluster.
Assuming the cluster is active most of the time, power is a proxy for this energy cost.

The second cost factor we study is the total memory capacity of the cluster.
Memory is a significant cost when building big data clusters.
For instance, memory is over 70\% of the cost of a Dell PowerEdge R930, if it is configured with the maximum memory~\cite{Dell2016}.

We present the results of our model under three different constraints by holding different cost-performance factors constant.
\begin{itemize}
 \item \emph{Performance provisioning}---Constrained response time (SLA): This represents designing a system to meet a performance requirement.
 \item \emph{Power provisioning}---Constrained power: This represents designing a system to fit a power envelope in a datacenter.
 \item \emph{Data capacity provisioning}---Constrained DRAM capacity: This represents designing a system around the data size of a workload.
\end{itemize}

\begin{table}
 \centering
 \begin{tabular}{| p{1.6cm} | c | c | c |}
  \hline
   & Traditional & Big-memory & Die-stacked \\ \hline \hline
  Number of blades & 800 & 1700 & 228 \\ \hline
  Number of chips & 3200 & 1700 & 1700 \\ \hline
  Cluster bandwidth & 320 TB/s & 320 TB/s & 384 TB/s \\ \hline
 \end{tabular}
 \caption{Requirements of a cluster of each system architecture given a 10~ms SLA.}
 \label{tab:10ms-cluster}
\end{table}

\subsection{Performance provisioning}

When designing a cluster to support an interactive big data workload, response time (or SLA, e.g., 99.9\% of queries must complete in 10~ms) is one possible constraining factor.
The SLA time includes many operations (e.g., multiple database queries, webpage rendering, and ad serving);
we focus on only one part of the entire response time: one query to an analytic database.
We look at three different SLA response times, 10~ms, 100~ms, and one second.
We present this spectrum of SLA response times since ``real-time'' may mean different response times to different providers.

\begin{figure*}[t]
 \begin{minipage}{.67\textwidth}
  \includegraphics[width=4.5in]{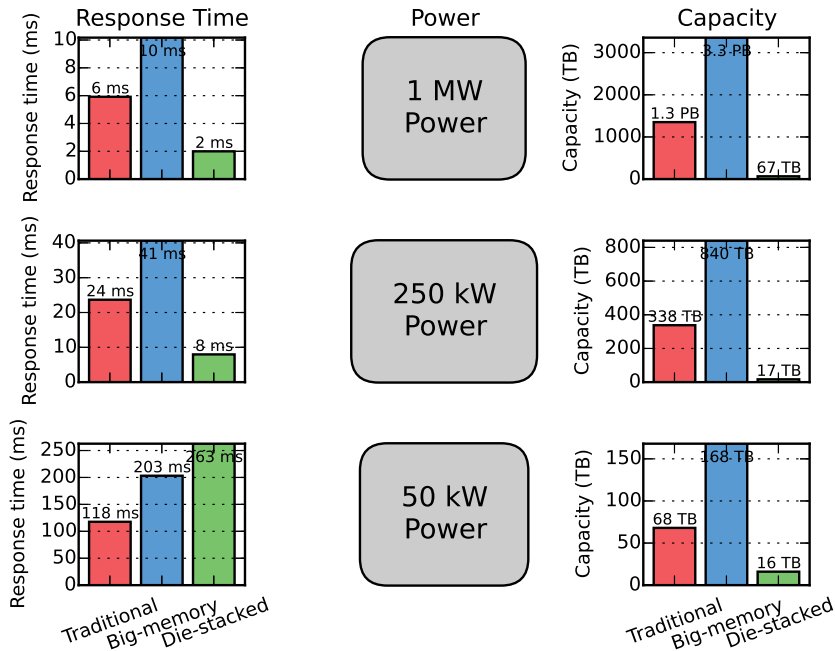}
 \end{minipage}
 \begin{minipage}{.33\textwidth}

  \begin{center}
  \fontsize{10pt}{10pt}\selectfont
  \underline{Power Provisioning Takeaways}
  \end{center}

  \textbf{Die-stacking is preferred with large and modest power budgets.}
  \begin{itemize}
   \item Die-stacked response time is $3\times$ and $5\times$ faster.
   \item Die-stacking performs 1.3--$2\times$ slower with tight power constraints.
  \end{itemize}
  \vspace{1em}

  \textbf{The big-memory system has the largest memory capacity.}
  \begin{itemize}
   \item The big-memory system has $3\times$ and $50\times$ more memory capacity.
   \item DRAM is a significant cost, but increased storage-only capacity may be beneficial.
  \end{itemize}

 \end{minipage}

 \caption{Response time and memory capacity of a cluster based on traditional, big-memory, and die-stacked servers with constant power envelope of either 1000~kW (1~MW), 250~kW, or 50~kW. Memory capacity is over provisioned if the power budget allows and capacity and performance are constrained by the allowed power. Each cluster is configured with multiple blades to exactly meet the power budget.}
 \label{fig:constant-power}
\end{figure*}

Similar to over provisioning disks to increase disk bandwidth and I/O operations per second, for the traditional and big-memory systems, we must over provision the total amount of main memory to increase the memory bandwidth.
To meet a certain SLA, the aggregate bandwidth of the cluster must be greater than the SLA response time divided by the amount of data accessed.
For instance, assuming a 10~ms SLA and a 3.2~TB working set, each system must support 320~TB/s of aggregate bandwidth.
For a cluster based on the traditional system to support 320~TB/s, it needs 800 blades and 3200 compute chips.
Table~\ref{tab:10ms-cluster} shows the requirements for all three systems given a 10~ms SLA.

Figure~\ref{fig:constant-perf} shows the power (Equation~\ref{equ:power}) and memory capacity (based on the required bandwidth to meet the SLA) for the clusters given a 10~ms SLA (top row), 100~ms SLA (second row), and one second SLA (bottom row).
This figure shows that if high performance (e.g., a 10~ms SLA) is required, the die-stacked system can provide significant benefits.
With a 10~ms SLA the die-stacked architecture uses almost $5\times$ less power, and does not need to be over provisioned at all.

The traditional and big-memory systems have significantly over provisioned memory capacity to be able to complete the workload in 10~ms.
This over provisioning means the traditional and big-memory platforms will be very expensive for this workload.
These platforms are over provisioned by a factor of $50\times$ and $213\times$, respectively.
These systems have low bandwidth compared to their memory capacity, and thus, require high memory capacity when performance is important.
However, the die-stacked system is the opposite: it has low capacity and high bandwidth, which is the most cost-effective system architecture when high performance is required.

Figure~\ref{fig:constant-perf} also shows the power and capacity for each system when the SLA is 100~ms and one second (second and third rows).
When the performance requirement is relaxed, the traditional and big-memory servers do not need to be as over provisioned, saving both power and memory cost.
In both of these cases, the die-stacked system uses about the same or more power than the current systems.
The die-stacked system uses more power because it requires more compute chips, more blades, and more memory modules to have the same capacity as the traditional and big-memory systems.
This increased number of chips puts the die-stacked system at a disadvantage when the SLA is less strict.

\begin{figure*}[t]
 \begin{minipage}{.67\textwidth}
  \includegraphics[width=4.5in]{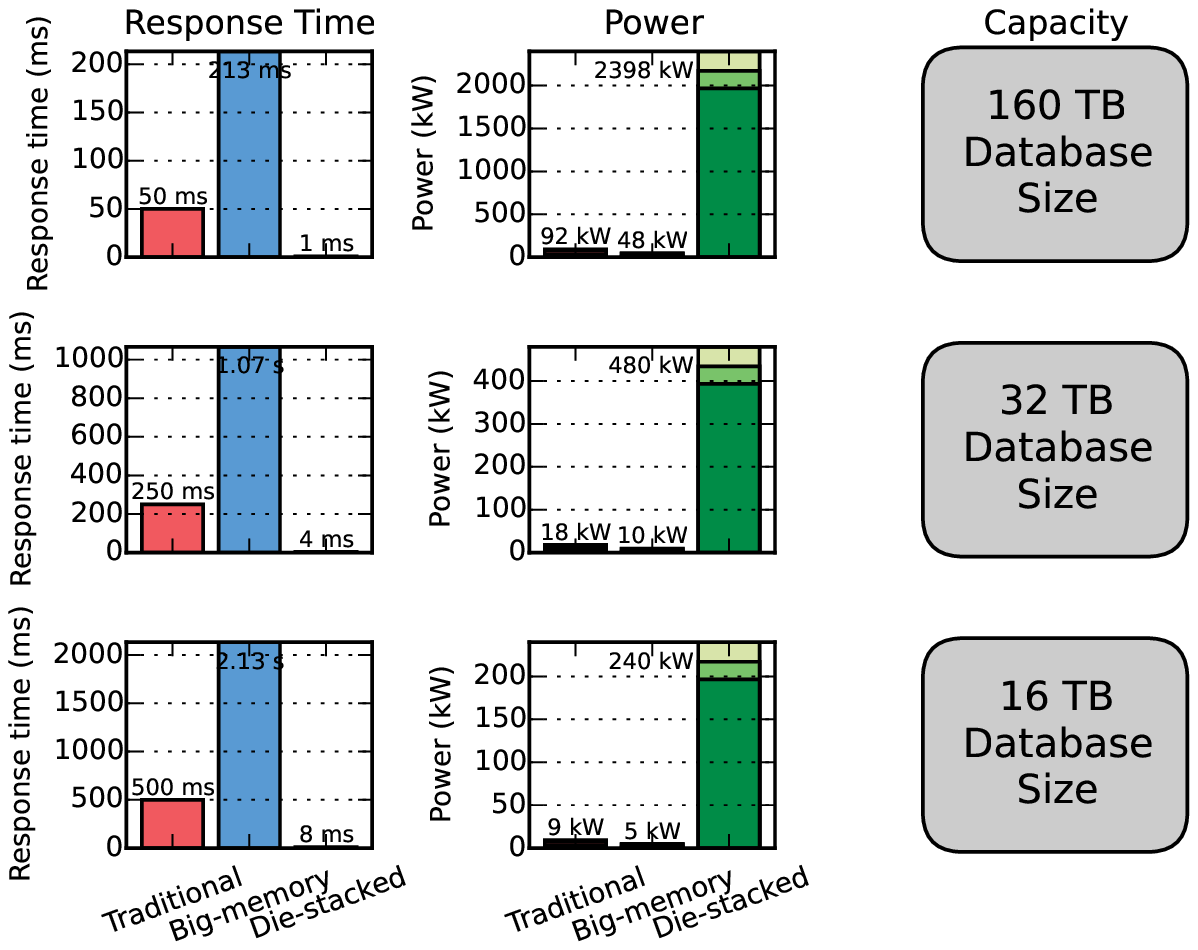}
 \end{minipage}
 \begin{minipage}{.33\textwidth}

  \begin{center}
  \fontsize{10pt}{10pt}\selectfont
  \underline{Data Capacity Provisioning Takeaways}
  \end{center}

  \textbf{Die-stacking greatly reduces response time.}
  \begin{itemize}
   \item Die-stacking reduces the response time by 60--256$\times$.
   \item Only die-stacking can meet a 10~ms SLA with these data capacities.
  \end{itemize}
  \vspace{1em}

  \textbf{Die-stacking uses much more power.}
  \begin{itemize}
   \item Die-stacking uses 25--50$\times$ more power.
   \item Die-stacking uses $5\times$ \emph{less} energy per query due to low response time.
  \end{itemize}

 \end{minipage}
 \caption{Response time and power of a cluster based on traditional, big-memory, and die-stacked servers with constant memory capacity of either 160~TB ($10\times$ data size), 32~TB ($2\times$ data size), or 16~TB ($1\times$ data size).}
 \label{fig:constant-capacity}
\end{figure*}

At an SLA of about 60~ms the power for the traditional and die-stacked systems is the same.
At this point, the only benefit to the die-stacked system is that it is less over provisioned, and could reduce the cost of extra DRAM, assuming die-stacked DRAM is the same cost as DDR4 DRAM.
As the complexity of the query increases (i.e., more data is accessed on each query) the crossover point moves to higher SLAs.
For instance, if the query touched 50\% of the data the crossover point is about 170~ms.
Also, as the density of the memory increases, the crossover point increases.
For instance, if the main memory was $8\times$ denser (e.g., using 64~Gb DRAM chips), the crossover point is about 800~ms.
Crossover points at higher SLAs imply the die-stacked system is cost-effective in more situations.

\subsection{Power provisioning}

Provisioning a datacenter for power distribution and removing the resulting heat is a large portion of the fixed cost of a datacenter~\cite{Barroso2013}.
To reduce the capitol investment overheads, it is important for the datacenter to use the amount of power provisioned for it.
Any provisioned power that is not used is wasted fixed cost.
Therefore, we investigate the tradeoffs given a fixed power budget.
We choose to hold the power of each system constant and one megawatt, 100~kW, and 50~kW.
As a point of reference, one megawatt is the power of a small datacenter~\cite{Barroso2013}.
We assume each system is used in a cluster that exactly meets the power budget (e.g., there are over 1300 traditional blades given 1~MW).

Figure~\ref{fig:constant-power} shows the performance (Equation~\ref{equ:response-time}) and memory capacity (based on the maximum number of blades for the given power budget) of each system architecture when provisioned to use 1~MW of power (top row), 100~kW (middle row), and 50~kW (bottom row).
At 1~MW each configuration can meet aggressive real-time SLA constraints of 10~ms.
However, the die-stacked architecture has a $5\times$ higher performance than the traditional and big-memory systems.
This decrease in response time translates into increased throughput in a real-world system.

The total memory capacity of each system given 1~MW of power is shown on the right side of Figure~\ref{fig:constant-power}.
Similar to Figure~\ref{fig:constant-perf}, the traditional and big-memory systems can support much higher capacity than the die-stacked system.
However, this increased capacity comes with an increased cost.

When the power limit is strict (e.g., 50~kW), the response time for the die-stacked system drops below the response time of the traditional and big-memory systems.
The reason for this behavior is that the die-stacked memory has a higher power per capacity (watts/bytes) due to its increased memory interface width and higher bandwidth.
Thus, when the power is strictly constrained, the die-stacked system is limited in the amount of compute power it can use.
In fact, for the 50~kW configuration, the die-stacked system only has enough power to use one core per compute chip.

\subsection{Data capacity provisioning}

Figure~\ref{fig:constant-capacity} shows the estimated performance (Equation~\ref{equ:response-time}) and power (Equation~\ref{equ:power}) using our model for three workload sizes: 160~TB (top row), 32~TB (middle row), and 16~TB (bottom row).
So far, all of the results has assumed a 16~TB workload size.
This data assumes that the workload only accesses 3.2~TB per query for each workload size.
Thus, for the larger workload sizes, the queries touch a smaller percentage of the total data.
If the complexity changed at the same rate as the capacity, all of the response times would be constant between different sizes.

This figure shows in all cases that the die-stacked system can perform a query $256\times$ faster than a big-memory machine and $60\times$ faster than a traditional server.
The die-stacked system improves performance for two reasons.
First, its aggregate bandwidth is much higher.
For instance, a die-stacked system with 16~TB has an aggregate bandwidth across all stacks of 512~TB/s.
The traditional server and big-memory servers have an aggregate bandwidth of 6.4~TB/s and 1.5~TB/s, respectively.
Second, to handle this increased bandwidth, the die-stacked system has many more compute chips, which in turn increases power (center column).
In fact, the die-stacked system uses 26--$50\times$ more power than the the traditional and big-memory systems.
This increased power usage significantly increases the cost of deploying a die-stacked cluster.

Interestingly, even though the die-stacked system has $340\times$ more bandwidth than the big-memory system, it is only $256\times$ faster.
This discrepancy is because there is not enough compute resources to keep up with the available bandwidth in the die-stacked system.
As compute chips become more capable or the density of main memory increases, the die-stacked system will increase the gap between other systems.

\begin{figure}
 \begin{subfigure}[c]{.45\columnwidth}
 \begin{center}
  \includegraphics[width=\textwidth]{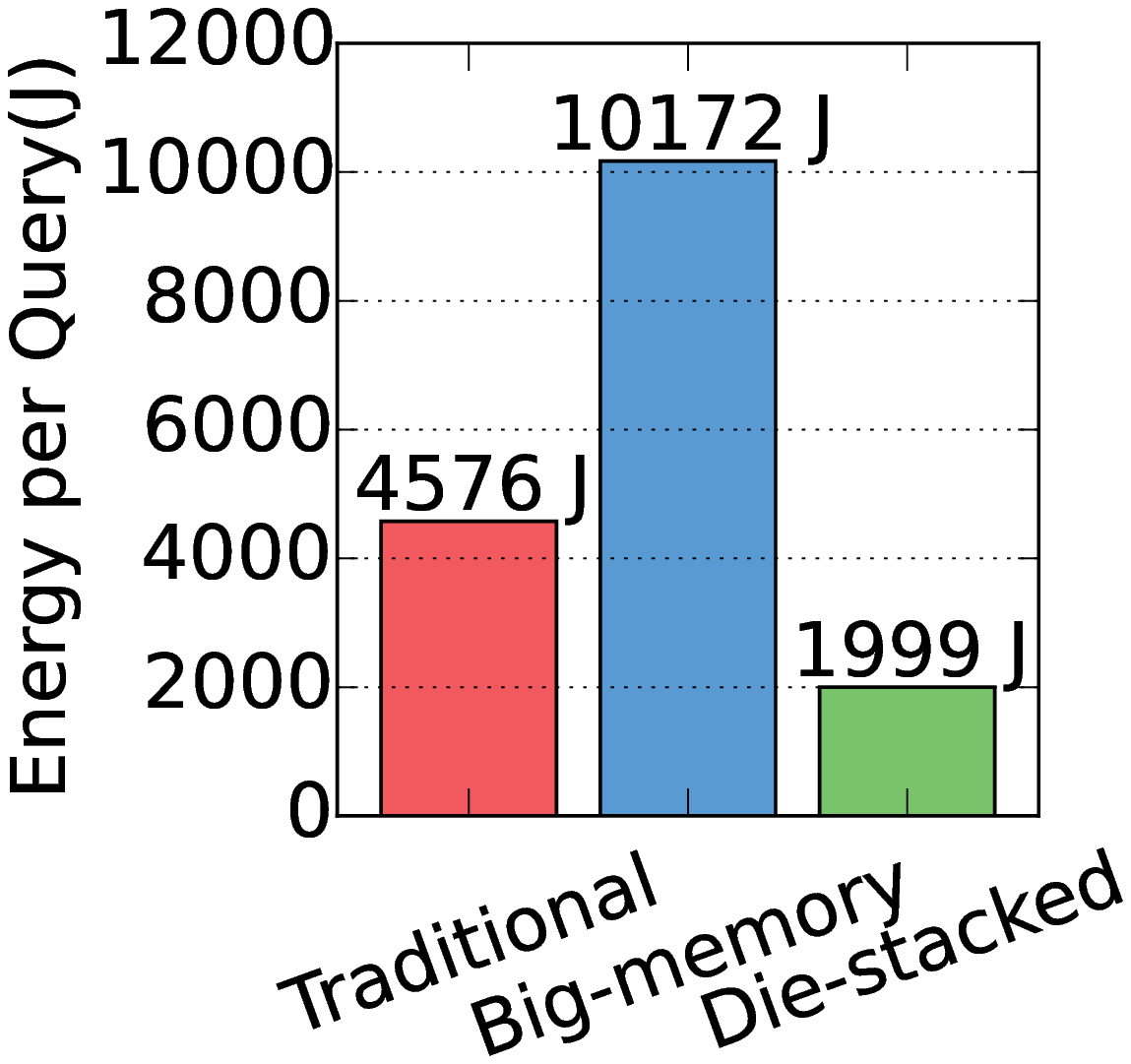}
  \caption{Energy for each system assuming each cluster is configured for a 16~TB database size (Figure~\ref{fig:constant-capacity} bottom row).}
  \label{fig:energy}
  \end{center}
 \end{subfigure}
 ~
 \begin{subfigure}[c]{.45\columnwidth}
 \begin{center}
  \includegraphics[width=\textwidth]{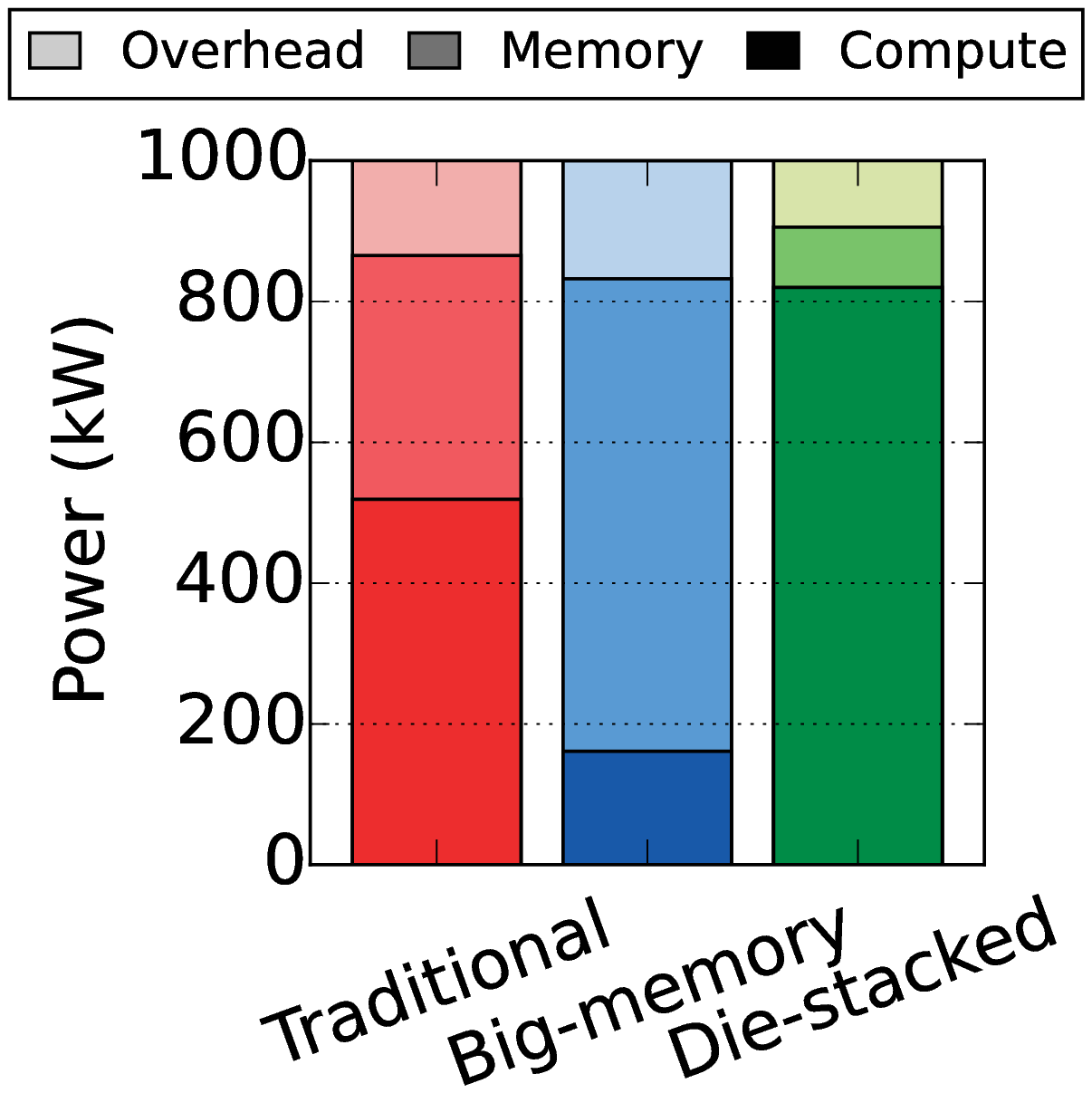}
  \caption{Breakdown of the components of power between compute power, memory power, and overhead power for a cluster with one megawatt of power.}
  \label{fig:power-breakdown}
  \end{center}
 \end{subfigure}
 \caption{Energy and power for each system.}
\end{figure}

Figure~\ref{fig:energy} shows the energy consumption of each system configured with a 16~TB database.
Even though the die-stacked system's power consumption is large, the die-stacked system is more energy efficient than either the traditional or big-memory system, using about $5\times$ less energy.
The die-stacked system does not need to use high-power chip-to-board interconnects or extra buffer-on-board chips.
The increased energy efficiency also comes from the fact that the die-stacked system ``races to halt''.
All of the extra energy from shared resources like the power supply, caches, disks, etc. is minimized since the die-stacked system completes the operation more quickly.
The relative energy efficiency of each system does not depend on the performance, power, or capacity provisioning.

\section{Discussion}
\label{sec:discussion}

In this section, we discuss the implications of the model results on designing a big data machine.
We first discuss the important areas for computer architects to focus to further increase the benefits of the future die-stacked architecture over the current server designs.
We then discuss some of the deficiencies of our analytical model.

\subsection{Improving system efficiency}

Figure~\ref{fig:power-breakdown} shows the breakdown of power between the compute, memory, and overhead for each system as a percentage of the whole system's power assuming a 1~MW power provisioned cluster.
The compute power is computed from Equation~\ref{equ:compute-power}, the memory power is computed from Equation~\ref{equ:mem-power}, and the overhead power assumes each blade (Equation~\ref{equ:blades}) uses an extra 100~W for peripheral components.
This figure shows that the traditional and big-memory servers' power is dominated by the memory chips.
However, the die-stacked server power is dominated by the compute chips.
None of the systems have a large portion of the power from the blade overhead component.

Figure~\ref{fig:power-breakdown} implies that architects should focus on reducing the compute energy in future die-stacked systems.
If the compute portion of power of the die-stacked system was reduced by $10\times$:
\begin{itemize}
 \item The die-stacked system would use less power when meeting high performance SLAs, but it does not fundamentally change the SLA tradeoffs.
 \item The capacity of the die-stacked system increases significantly when you have a fixed power budget of more than 100~kW.
 \item The die-stacked system has reduced power in the case of fixed data capacity, but it does not fundamentally change the response time tradeoffs.
 \item The die-stacked system would have increased energy efficiency.
\end{itemize}

Alternatively, architects could concentrate on increasing the main memory density.
One promising technology that increases density over DRAM is non-volatile memories~\cite{Song2013}.
Increasing the density of main memory does not directly affect the performance in any system architecture.
However, increased density does reduce the required power (especially for the die-stacked system) as fewer chips and fewer blades are required to meet the capacity requirement.
Increasing the density by $8\times$ affects our results in the following ways:
\begin{itemize}
 \item The die-stacked architecture becomes cost-effective at higher SLAs, but it is still less cost-effective than the traditional system with a one second SLA.
 \item With fixed power provisioning, increasing the density does not change the performance, but does increase the capacity for all systems.
 \item For a fixed data capacity, increasing the density decreases the power required for all systems and makes the performance worse for the traditional and big-memory systems due to the lower bandwidth-capacity ratio.
\end{itemize}

\subsection{Model deficiencies}

In this work, we use a high-level model of the performance and power of a big data cluster.
There are many important factors that we ignore.
First, we only focus on one potential big data workload.
However, we believe our results generalize to any bandwidth-bound workload.
Second, the inputs we use for our model significantly affect the results.
We use data from manufacturers and datasheets.
However, it is likely these numbers vary between providers and will change over time.
For these reasons, we have released our model as an interactive online application for readers to choose the numbers which are best for their systems. See \url{https://research.cs.wisc.edu/multifacet/bpoe16_3d_bandwidth_model/}.

Also, we do not know whether the compute energy is due to data movement or actual computation.
Since we use real hardware to measure the energy of a computation, we cannot break down the compute energy into its components.
More deeply understanding the power and energy implications in die-stacked systems is interesting future work.

In this work, we assume that each processor only accesses its local memory and ignored the time required for communication between processors in the cluster.
Even with regular data-parallel workloads like those we evaluated, there will be some amount of communication between processors.
Largely, these overheads will be the same for each of the systems we evaluated.
However, it is possible that since the die-stacked system has less data per processor that the
communication overheads will be larger than for systems with more data per processor.
Our model does not consider the communication between processors.

Finally, our model only applies to workloads that are limited by memory bandwidth.
We defined workload complexity in terms of the data required for the computation.
However, another common measure of complexity is ``arithmetic intensity'': the ratio of compute operations (e.g., FLOPS) to memory operations.
Our model does not directly take this characteristic into consideration.
Current big data workloads do few compute operations per byte of data, which causes these workloads to be memory-bandwidth bound.
Additionally, as more big data applications take advantage of vector extensions like SIMD their arithmetic intensity will be even lower putting more pressure on the memory bandwidth.

\section{Conclusions}
\label{sec:conclusions}

We investigated how the system architecture affects the power and memory capacity cost-performance tradeoff for three possible designs, a traditional server, a big-memory server, and a novel die-stacked server for one big data workload.
We find the traditional server and the big-memory server are not architected for low-latency big data workloads.
These current systems have a poor ratio of memory bandwidth to memory capacity.
To increase the performance of big data workloads, these systems need a larger bandwidth-capacity ratio, increasing the memory bandwidth without increasing the memory capacity.

On the other hand, the die-stacked system has a bandwidth-capacity ratio that is too large to be cost-effective in many situations.
The die-stacked system provides greatly increased bandwidth, but to the detriment of memory capacity.
In fact, we needed over 2000 stacks to meet the same capacity as eight big-memory systems.
Because the die-stacked system requires so many stacks to reach a high capacity, the power requirements may be a cost-limiting factor.

We find the die-stacked architecture of today is cost-effective with complex queries that touch a significant fraction of the total data and there is a tight (10~ms) SLA.
Looking forward, architects should focus on reducing the compute power in these die-stacked systems or increasing the memory density.
These changes will allow the die-stacked architecture to be cost-effective in more situations.

\acks
We thank our anonymous reviewers, Jignesh Patel, and members of the Multifacet group for their insightful comments and feedback on the paper.
This work is supported in part by the National Science Foundation (CCF-1218323, CNS-1302260, CCF-1438992, CCF-1533885), Cisco Systems Distinguished Graduate Fellowship, John P. Morgridge Chair, Google, and the University of Wisconsin--Madison (Named Professorship).
Hill and Wood have a significant financial interest in AMD and Google.


\bibliographystyle{abbrvnat}
\bibliography{adbm}


\end{document}